# Scaling Two-Dimensional Semiconductor Nanoribbons for High-Performance Electronics

Hao-Yu Lan,[1,2] Shao-Heng Yang,[1,2] Yongjae Cho,[1,2,6] Yuanqiu Tan,[1,2] Jun Cai,[1,2] Zheng Sun,[1,2] Chenyang Li,[4,5] Lin-Yun Huang,[5] Yi Wan,[4] Lain-Jong Li,[4] Thomas Beechem,[2,3] Joerg Appenzeller,[1,2] and Zhihong Chen[1,2]

[1]*Electrical and Computer Engineering, Purdue University, West Lafayette, Indiana, USA*

[2]*Birck Nanotechnology Center, Purdue University, West Lafayette, Indiana, USA*

[3]*Mechanical Engineering, Purdue University, West Lafayette, Indiana, USA*

[4]*Materials Science and Engineering, National University of Singapore, Singapore*

[5]*Nexstrom Pte. Ltd., Singapore, Singapore*

[6]*Department of Physics, Dankook University, Cheonan, Chungcheongnam-do, South Korea*

**To whom correspondence should be addressed.*

*E-mail: zhchen@purdue.edu*

## Abstract

As silicon transistors scale toward future technology nodes, three-dimensional architectures—including gate-all-around (GAA) nanoribbon and complementary field-effect transistors (CFETs)—require channel widths in the tens of nanometers to meet density targets. Monolayer transition metal dichalcogenides (TMDs), with their atomically thin bodies, are promising channel materials for these architectures, yet most TMD-based FETs remain limited to micrometer-scale widths. Here, we show that channel width scaling of monolayer $MoS_2$ nanoribbon transistors not only preserves but also enhances device performance. Reducing the channel width from hundreds of nanometers to ~30-40 nm increases the median on-current density by ~42% and reduces the median subthreshold swing by ~16%, with a champion device reaching 995 μA μm$^{-1}$ at a drain-to-source voltage of 1 V and an overdrive voltage of 2.5 V. We attribute these improvements to three mechanisms: minimal edge-induced disorder, enhanced gate electrostatics at ribbon edges, and more efficient side-contact injection, together reducing contact resistance from ~860 Ω μm to ~270 Ω μm. Extending the platform to *n*-type $WS_2$ and

*p*-type $WSe_2$ FETs, we achieve $WSe_2$ p-FET on-currents of 357 μA $\mu m^{-1}$. These findings suggest that monolayer TMD nanoribbon FETs are promising candidates for future ultra-scaled electronics.

## Introduction

Silicon transistor scaling is a cornerstone of modern electronics, enabling continued improvements in power efficiency, performance, and integration density over the past five decades. As gate lengths have approached the nanometer scale, planar transistor structures have evolved into three-dimensional architectures, including FinFETs[1], GAA nanosheet FETs (NS-FETs)[2,3], nanoribbon FETs (RibbonFETs)[4], and vertically stacked complementary FETs (CFETs)[5], which offer improved electrostatics and scalability. In these architectures, nanoribbon channels with widths in the tens of nanometers are required to meet density targets and minimize device footprint. However, continued scaling of the silicon channel is constrained by fundamental physical limits, driving the search for alternative channel materials. Monolayer transition metal dichalcogenides (TMDs), with their atomically thin bodies and naturally passivated surfaces, provide a promising platform for next-generation ultra-scaled transistors[6–13]. Although sub-100 nm channel-lengths with competitive performance have been demonstrated[12,14–27], most devices still remain limited to micrometer-scale channel widths, far from the tens-of-nanometer dimensions required for practical GAA integration.

To meet stringent density and performance requirements at future technology nodes, continued scaling of GAA transistors demands further reduction of nanoribbon channel width, potentially below 10 nm[28], as projected by the International Roadmap for Devices and Systems (IRDS)[29]. These projections establish channel width as a critical design parameter alongside gate length and oxide thickness. Despite this clear technological imperative, it remains unknown whether aggressive width scaling in monolayer TMDs preserves or degrades key device metrics—including on-current density, threshold voltage, subthreshold swing, and contact resistance—and which physical mechanisms govern this behavior.

Here, we demonstrate that aggressive width scaling in monolayer TMD nanoribbon transistors not only preserves but also enhances device performance within the experimentally accessible regime. Scaling the channel width from hundreds of nanometers to ~30-40 nm yields a ~42% increase in median on-current, a champion-device maximum current density of ~995 μA $\mu m^{-1}$, and a minimum subthreshold swing of 75 mV $dec^{-1}$. Through statistical electrical analysis, supported by optical characterization and TCAD simulations, we identify the physical origin of this enhancement. Contrary to conventional expectations, width scaling introduces minimal edge-induced disorder—preserving material quality and off-state behavior—while enhancing gate electrostatics and contact injection. Narrower ribbons intensify the gate electric field and carrier density at the ribbon edges while enabling more efficient side-contact injection, together reducing contact resistance from ~860 to ~270 Ω μm. Our nanoribbon devices surpass prior reports on monolayer TMD nanoribbon transistors at comparable channel dimensions. Beyond $MoS_2$, we extend the nanoribbon platform to high-performance *n*-type $WS_2$ and *p*-type $WSe_2$ nanoribbon FETs, and demonstrate scalability in $MoS_2$ nanoribbon transistor arrays with a contact length of ~30 nm, a channel length of ~30 nm, and an equivalent oxide thickness (EOT) of ~0.9 nm. These results establish width scaling as an effective and underexplored design parameter for future GAA and CFET transistor architectures.

## Results

### Scaling 2D monolayer $MoS_2$ nanoribbon for high-performance transistors

Compared to bulk silicon transistors, 2D semiconductor nanoribbon FETs offer distinct structural advantages, as illustrated in **Fig. 1a-b**. The GAA CFET architecture enabled by 2D nanoribbon channels (**Fig. 1a**) benefits from the atomically thin body of monolayer TMDs, which not only allows ultimate length scaling but also confines edge states exclusively to the ribbon edges. Unlike bulk silicon—which exhibits surface states in the thickness dimension and edge states in the lateral dimension—2D nanoribbons have naturally passivated top and bottom surfaces ideally with no dangling bonds, minimizing edge-state-induced performance degradation (**Fig. 1b**). While graphene nanoribbons suffer significant edge-related mobility degradation upon width scaling[30], the corresponding impacts on 2D TMD-based FETs remain less understood, motivating a systematic investigation of width-scaling effects.

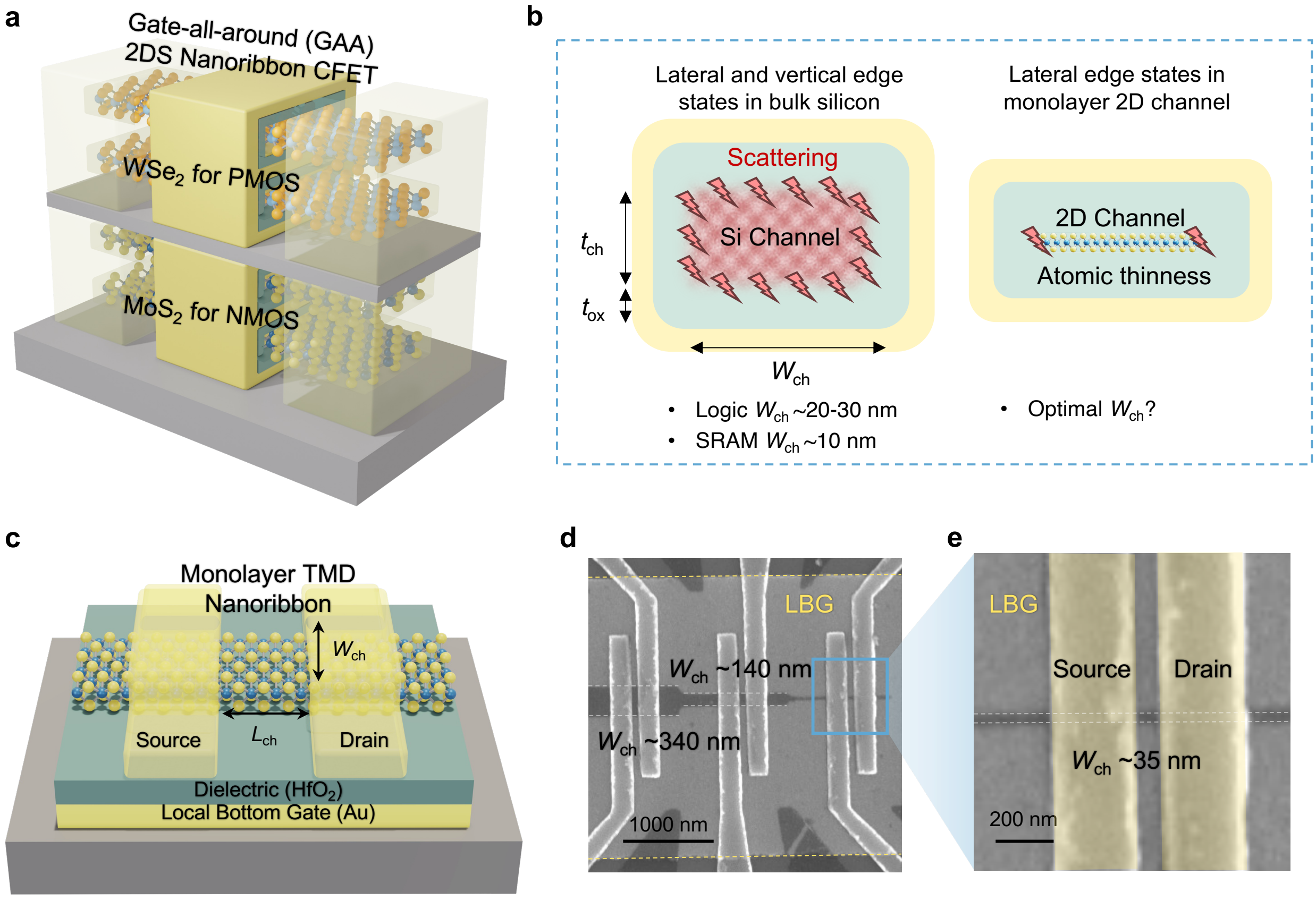

**Figure 1. Scaled monolayer $MoS_2$ nanoribbon FETs. (a)** Schematic of gate-all-around (GAA) complementary field-effect transistors (CFETs) enabled by 2D semiconductor nanoribbon channels, with $WSe_2$ as the *p*-type channel and $MoS_2$ as the *n*-type channel. **(b)** Conceptual comparison of edge states in bulk silicon GAA versus monolayer 2D nanoribbon channels, illustrating reduced edge-state density in 2D materials owing to naturally passivated surfaces in the vertical dimension. **(c)** Device schematic of a monolayer $MoS_2$ nanoribbon transistor with Cr/Au local bottom gate (LBG), 3 nm $HfO_2$ gate dielectric, and Ni source/drain contacts, with channel width scaled to ~35 nm. **(d)** Top-view SEM image of fabricated monolayer $MoS_2$ nanoribbon transistors with varying channel widths. **(e)** Zoomed-in SEM image of a representative device confirming a channel width ($W_{ch}$) of ~35 nm and channel length ($L_{ch}$) of ~75 nm.

**Figure 1c** shows a device schematic of the monolayer $MoS_2$ nanoribbon FETs studied here as a model system, featuring a Cr/Au local bottom gate (LBG), 3 nm $HfO_2$ gate dielectric grown at 90 °C, Ni source/drain contacts, and channel widths scaled to ~30-40 nm. The ultrathin dielectric enables channel-length scaling, reduces device-to-device variability, and mitigates the influence of Schottky barrier height at the metal-semiconductor interface[17]. Detailed fabrication procedures, including the $Cl_2/O_2$ plasma etching recipe used for nanoribbon patterning, are described in the **Methods** section. In our process flow, nanoribbons are patterned prior to contact formation (**Supplementary Fig. 1)**. For widths down to ~35 nm, no noticeable degradation in adhesion or structural integrity compared with wider (microribbon) devices is observed, nor is delamination observed during subsequent processing. The overall fabrication yield exceeds 90%, with non-functional devices attributed primarily to gate leakage through the ultrathin dielectric or intrinsic channel defects, rather than to nanoribbon delamination.

**Figure 1d** shows a top-view scanning electron microscopy (SEM) image of fabricated $MoS_2$ nanoribbon devices with varying widths; a zoomed-in image (**Fig. 1e**) confirms $W_{ch}$ ~35 nm and $L_{ch}$ ~75 nm. Additional SEM images are provided in **Supplementary Fig. 2**. The nanoribbon channel width ($W_{ch}$) is extracted from SEM images using contrast line profiles perpendicular to the width direction; the width is defined as the full-width at half-maximum (FWHM) from Gaussian fitting (**Supplementary Fig. 2**). This method provides a consistent and reproducible width definition with an estimated uncertainty of ~3-5 nm. AFM characterization (**Supplementary Fig. 3**) further confirms the nanoscale channel width down to ~30 nm and structural continuity of the nanoribbon. Raman spectroscopy confirming preserved $MoS_2$ crystallinity following nanoribbon patterning is presented and discussed in the following section.

## Channel width scaling of monolayer TMD nanoribbon FETs

To investigate the effects of width scaling on monolayer $MoS_2$ nanoribbon transistors, we fabricated sets of devices with varying widths and a channel length of $L_{ch}$ = 55-75 nm. **Figure 2a-b** presents representative log-scale and linear-scale transfer characteristics ($I_D$-$V_{GS}$) for two channel widths ($W_{ch}$ = 540 nm and $W_{ch}$ = 35 nm). Scaling of the channel width results in enhanced transistor performance in both the on-state and off-state regimes, achieving a high maximum current density ($I_{max}$) of ~727 μA μm$^{-1}$ at a drain-to-source voltage ($V_{DS}$) of 1 V and an overdrive voltage ($V_{OV} = V_{GS} - V_{T,lin}$) of 1.65 V (**Fig. 2a-b**), where $V_{GS}$ is the gate-to-source voltage and $V_{T,lin}$ is extracted by linear extrapolation at the point of maximum transconductance. In the deep off-state, the measured drain current is leakage-dominated; when normalized by width (μA μm$^{-1}$), similar absolute leakage yields a larger current density in narrower ribbons. No increase in absolute leakage is observed with decreasing width.

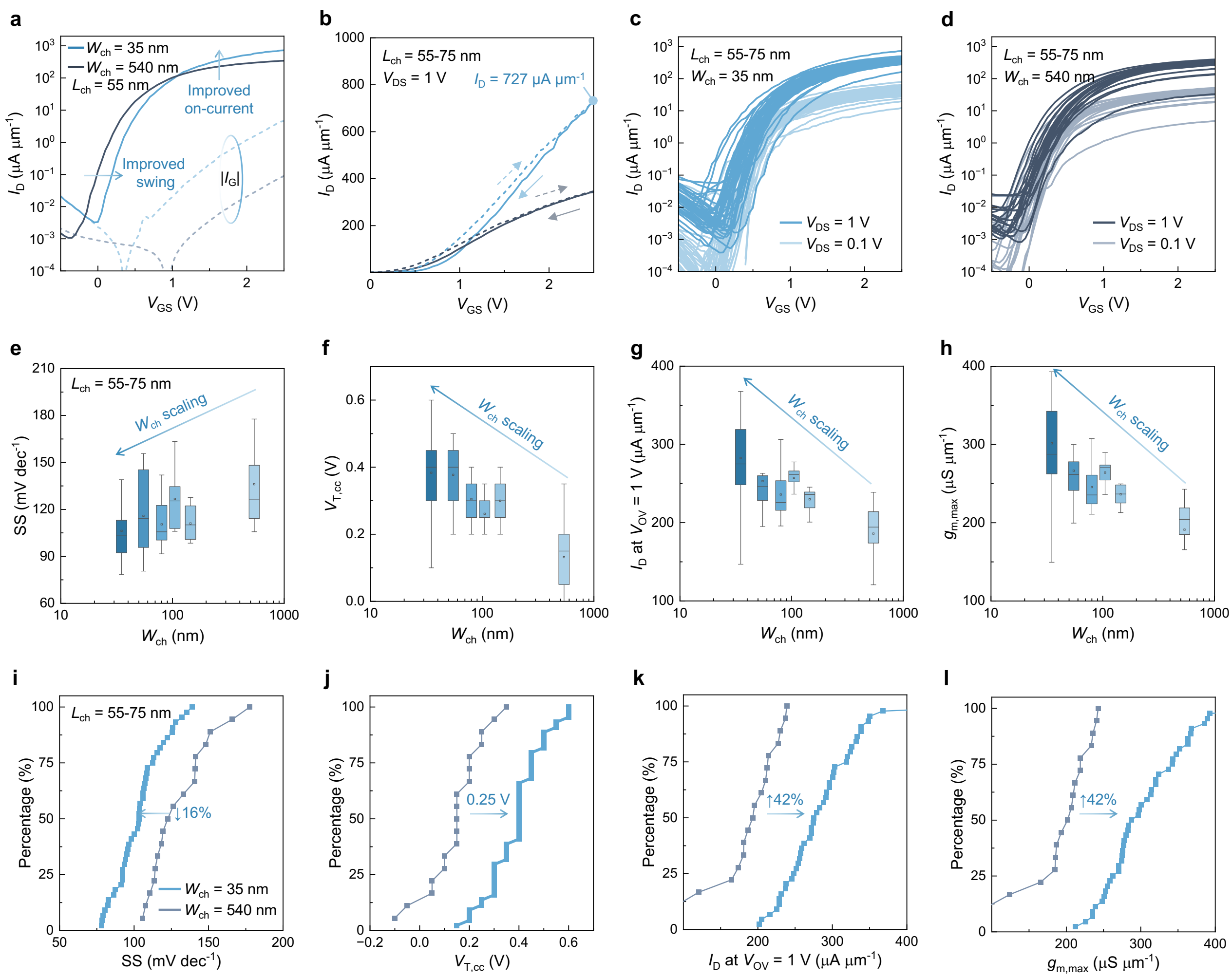

**Figure 2. Channel width scaling of monolayer $MoS_2$ nanoribbon FETs with performance enhancement.** **(a)** Log-scale and **(b)** linear-scale transfer characteristics ($I_D$-$V_{GS}$) for representative devices with channel widths of 540 nm and 35 nm, showing enhanced on-state and off-state performance upon width scaling. The 35-nm width device achieves maximum current density ($I_{max}$) ~727 μA μm$^{-1}$ at $V_{DS}$ = 1 V and $V_{OV}$ = 1.65 V. **(c)-(d)** $I_D$-$V_{GS}$ for multiple devices with narrow ($W_{ch}$ = 35 nm, **c**) and wide ($W_{ch}$ = 540 nm, **d**) channel widths, demonstrating reproducible performance. **(e)-(h)**, Statistical box plots of subthreshold swing (SS), threshold voltage ($V_{T,cc}$), on-current ($I_{on}$), and peak transconductance ($g_{m,max}$) as a function of channel width ($W_{ch}$), showing systematic performance improvement with scaling. $I_{on}$ is extracted at $V_{DS}$ = 1 V and $V_{OV}$ = 1 V. **(i)–(l)**, Cumulative distribution functions (CDFs) of *SS*, $V_{T,cc}$, $I_{on}$, and $g_{m,max}$ comparing normal-width (540 nm) and scaled (35 nm) nanoribbon devices. Scaled devices show improved median SS (from 123 mV dec$^{-1}$ to 103 mV dec$^{-1}$), a positive $V_{T,cc}$ shift (~0.25 V), a ~42% increase in median $I_{on}$ (from 193 μA μm$^{-1}$ to 275 μA μm$^{-1}$), and a ~42% increase in median $g_{m,max}$.

A positive shift in threshold voltage ($V_{T,cc}$) and improved subthreshold swing (SS) are observed in the narrow nanoribbon (**Fig. 2a**), indicating improved gate electrostatics. Note that the SS was determined from the inverse of the minimum slope of the log-scale transfer curve at $V_{DS}$ = 0.1 V, and $V_{T,cc}$ was defined at a constant drain current of 100 nA μm$^{-1}$ at $V_{DS}$ = 0.1 V for off-state evaluation[31]. Such enhanced gate control through width scaling has previously been reported for oxide semiconductors[32], and this study represents the first observation in 2D monolayer $MoS_2$ devices, in contrast to prior studies in which width scaling typically degraded device performance[33–37]. The raw transfer characteristics of multiple short-channel devices with channel length ($L_{ch}$ = 55-75 nm) at each width are shown in **Fig. 2c-d**, confirming reproducible performance across devices for both narrow ($W_{ch}$ = 35 nm, **Fig. 2c**) and wide ($W_{ch}$ = 540 nm, **Fig. 2d**) channel widths. The individual transfer curves for the full range of channel widths studied ($W_{ch}$ = 35, 55, 80, 105, 145, and 540 nm) are shown in **Supplementary Fig. 4**.

To systematically quantify the impact of width scaling, we extracted key device metrics (SS, $V_{T,cc}$, $I_{on}$, $g_{m,max}$) from hundreds of devices and plotted them as a function of channel width (**Fig. 2e-h**). For a fair comparison of on-state performance, all devices were evaluated for on-current ($I_{on}$) at the same overdrive voltage ($V_{OV} = V_{GS} - V_{T,lin}$) of 1 V. A clear improvement trend in all metrics with decreasing $W_{ch}$ demonstrates the consistent benefit of width scaling. In particular, the relationship between $V_{T,cc}$ and SS (**Fig. 2e,f**) reveals that narrower devices exhibit more positive $V_{T,cc}$ and lower SS, suggesting stronger electrostatic control in the narrower channels. As displayed in **Fig. 2g,h**, the improvement in $I_{on}$ and $g_{m,max}$ with decreasing $W_{ch}$ in short-channel devices (contact-limited regime) suggests that, in addition to stronger electrostatic control, there is additional contact-related enhancement, as discussed in the following section.

**Figures 2i-l** present cumulative distribution functions (CDFs) of SS, $V_{T,cc}$, $I_{on}$, and $g_{m,max}$ for normal-width (540 nm) and aggressively scaled (35 nm) devices. The 35-nm width devices show an improvement in median SS from 123 mV dec$^{-1}$ to 103 mV dec$^{-1}$ and a positive $V_{T,cc}$ shift of ~0.25 V, alongside a ~42% increase in median $I_{on}$ (from 193 μA μm$^{-1}$ to 275 μA μm$^{-1}$) and a ~42% increase in median $g_{m,max}$. An increase in on-state variability ($I_{on}$ and $g_{m,max}$) shown in **Fig. 2g,h** and **Fig. 2k,l** is observed in short-channel devices (contact-limited regime), which we attribute mainly to Schottky barrier height inhomogeneity at the Ni/$MoS_2$ interface, consistent with previously reported results on $MoS_2$ devices[38]. Understanding how Schottky barrier height inhomogeneity scales with channel width will be key to reducing device variability.

## Uncovering the origins of performance enhancement in nanoribbon transistors

To understand the mechanisms driving performance enhancement in monolayer $MoS_2$ nanoribbon FETs, we first confirmed that nanoribbon patterning preserves the intrinsic material quality. Raman spectroscopy of patterned nanoribbons (**Fig. 3a**) shows no detectable broadening in the characteristic E' and $A_1$' peaks of $MoS_2$, confirming preserved crystallinity following the $Cl_2/O_2$ plasma etching process. The inset SEM image in **Fig. 3a** shows the nanoribbon array used for this measurement. To probe the charge state at the nanoribbon edge, we performed spatially resolved photoluminescence (PL) characterization (**Fig. 3b-c**). The PL intensity map and K-means clustering (**Fig. 3b**) spatially separate bulk and edge regions within a nanoribbon device. The corresponding PL spectra (**Fig. 3c**) reveal a notable positive shift (~ 5 meV) in exciton energy at the nanoribbon edge compared to the bulk region, consistent with reduced electron density, suggesting partial depletion at the edge[39]. **Supplementary Fig. 5** shows the fitted peak position acquired using 532 nm laser on a natural and lithographically patterned $MoS_2$ edge. In all cases, the exciton energy increases upon moving across the edge. The magnitude of this change is equal irrespective of the type of patterning, implying that lithographic patterning and etching does not significantly damage or alter the intrinsic properties of $MoS_2$ edges.

Interestingly, lithographically patterned edges exhibit slightly higher exciton energy compared to naturally formed edges. We hypothesize this subtle difference arises from near-edge oxygen incorporation introduced during the $Cl_2/O_2$ plasma etching process. This localized oxygen incorporation could passivate sulfur vacancies through partial oxidation, inducing mild depletion—consistent with our observed PL shifts and supported by prior literature. The presence of oxygen passivation effectively removes mid-gap states originating from vacancies, thereby potentially reducing the edge-related mobility degradation. Such an effect aligns closely with previously reported improvements observed in oxygen-doped $MoS_2$ synthesized via direct methods[40] and oxygen-incorporated $MoS_2$ from chemical vapor deposition (CVD)[41].

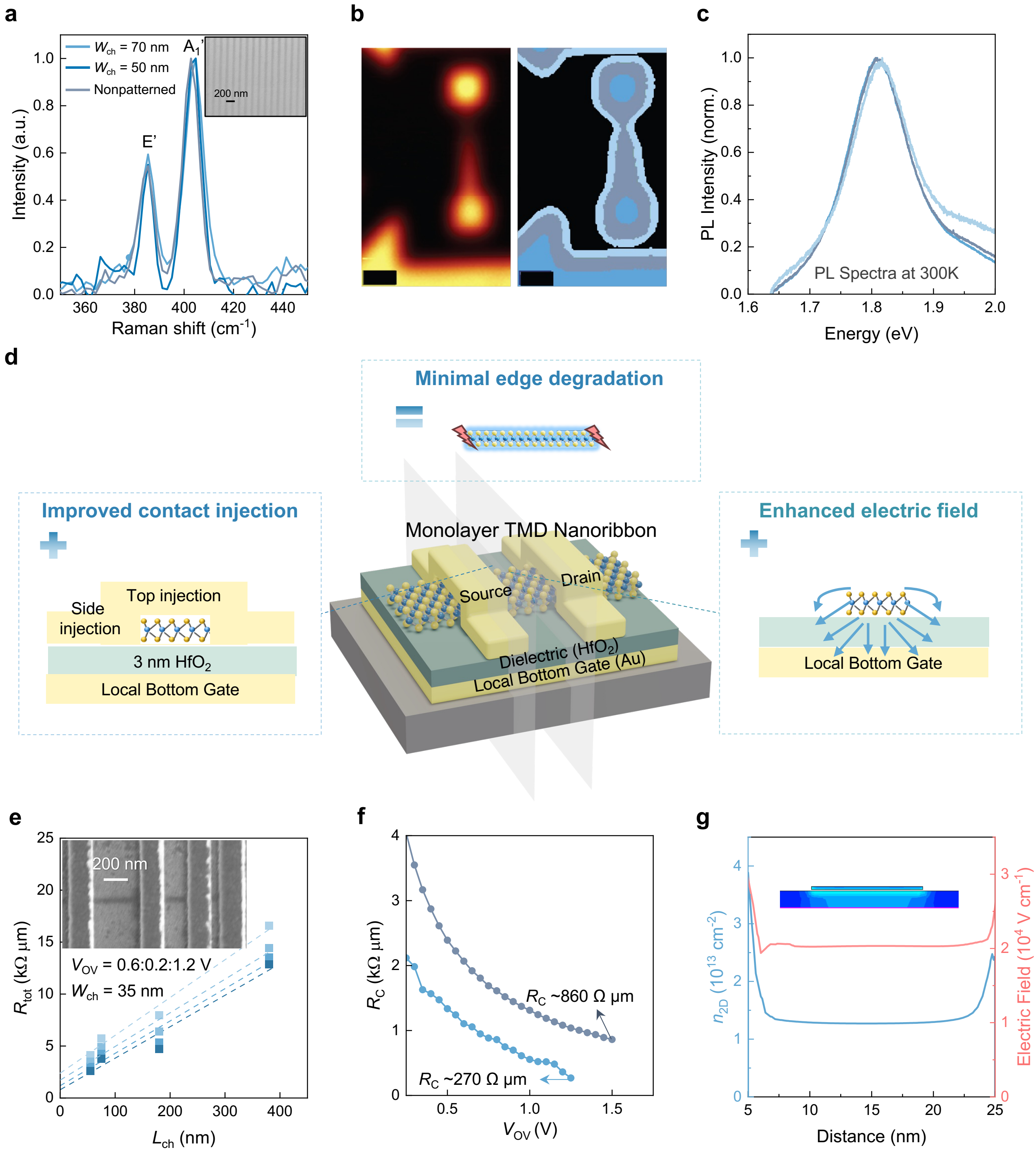

**Figure 3. Physical origin of performance enhancement in monolayer $MoS_2$ nanoribbon FETs. (a)** Raman spectra of patterned and unpatterned $MoS_2$, confirming preserved material crystallinity after nanoribbon fabrication, with characteristic E' and $A_1$' peaks showing negligible shift or broadening. Inset: SEM image of the nanoribbon array used for measurement. **(b)** Spatially resolved photoluminescence (PL) intensity map and corresponding K-means clustering, separating bulk and edge regions within a nanoribbon device. Scale bar, 500 nm. **(c)** Room-temperature PL spectra comparing bulk and edge emission, showing a positive exciton energy shift at the nanoribbon edge consistent with minimal edge-induced disorder and preserved optical quality. **(d)** Device schematic summarizing the key mechanisms of performance enhancement: improved contact injection

from the nanoribbon edge (left), and enhanced gate electric field from the local bottom gate (right). **(e)** Total resistance ($R_{tot}$) as a function of $L_{ch}$ at $V_{OV}$ = 0.6-1.2 V, extracted from TLM measurements. Inset: SEM image of the fabricated nanoribbon TLM array. **(f)** Contact resistance ($R_C$) as a function of $V_{OV}$ for wide (540 nm, dark blue) and narrow (35 nm, blue) nanoribbon devices, showing substantially reduced $R_C$ in nanoribbon devices. **(g)** TCAD-simulated 2D electric field and electron density distributions along the channel width in the on-state ($V_{DS}$ = 1 V, $V_{GS}$ = 3 V), showing intensified edge fields and increased carrier density at nanoribbon edges.

The physical mechanisms driving this performance enhancement are summarized in **Fig. 3d**: width scaling enhances contact injection at the nanoribbon edges by increasing the contribution of side contacts (left), while increasing the local electric field from the bottom gate in narrow channels (right). To quantify these effects electrically, we first compare contact resistance for wide (540 nm) and narrow (35 nm) nanoribbon devices. **Fig. 3e** shows the total resistance ($R_{tot}$) as a function of channel length at overdrive voltages ($V_{OV}$ = 0.6 V to 1.2 V), extracted from one representative transfer length method (TLM) measurement (inset SEM in **Fig. 3e**). **Figure 3f** shows contact resistance ($R_C$) as a function of $V_{OV}$ for wide (540 nm, dark blue) and narrow (35 nm, blue) nanoribbon devices. The 35-nm-width nanoribbon devices exhibit substantially reduced $R_C$, approaching ~270 Ω μm at $V_{OV}$ = 1.25 V, compared to ~860 Ω μm at $V_{OV}$ = 1.5 V for wide devices, consistent with enhanced contact gating and enhanced side contact injection. Statistical distributions of $R_C$, sheet resistance ($R_{sh}$), and effective mobility ($\mu_{eff}$) across multiple TLM structures are provided in **Supplementary Fig. 6**.

To understand the electrostatic enhancement from nanoribbon devices, Sentaurus TCAD simulations were performed (**Fig. 3g**). The simulated two-dimensional carrier density ($n_{2D}$) and lateral electric field distributions along the channel width clearly reveal intensified edge fields and elevated carrier densities near the nanoribbon edges in the on-state ($V_{DS}$ = 1 V, $V_{GS}$ = 3 V). Collectively, these findings confirm that minimal edge disorder, improved electrostatics, and enhanced side-contact injection are the three key mechanisms driving performance enhancement in aggressively scaled $MoS_2$ nanoribbon devices.

To quantify the relative contributions of these mechanisms, we compare the on-current improvement between long-channel (channel-resistance-dominated) and short-channel (contact-resistance-dominated) devices. Specifically, long-channel devices show a median on-current improvement of ~11% (**Supplementary Fig. 7**), consistent with ~10% enhanced electrostatics from TCAD simulations. In contrast, short-channel devices (contact-limited) exhibit a larger enhancement of ~42%, which we attribute to both enhanced electrostatics and improved contact injection from the nanoribbon edges. This enhanced side-injection mechanism of the nanoribbon is further supported by our IEDM 2025 proceeding[42], which demonstrates that side-contact injections alone can sustain high on-current.

## High-performance complementary nanoribbon transistor based on monolayer TMDs

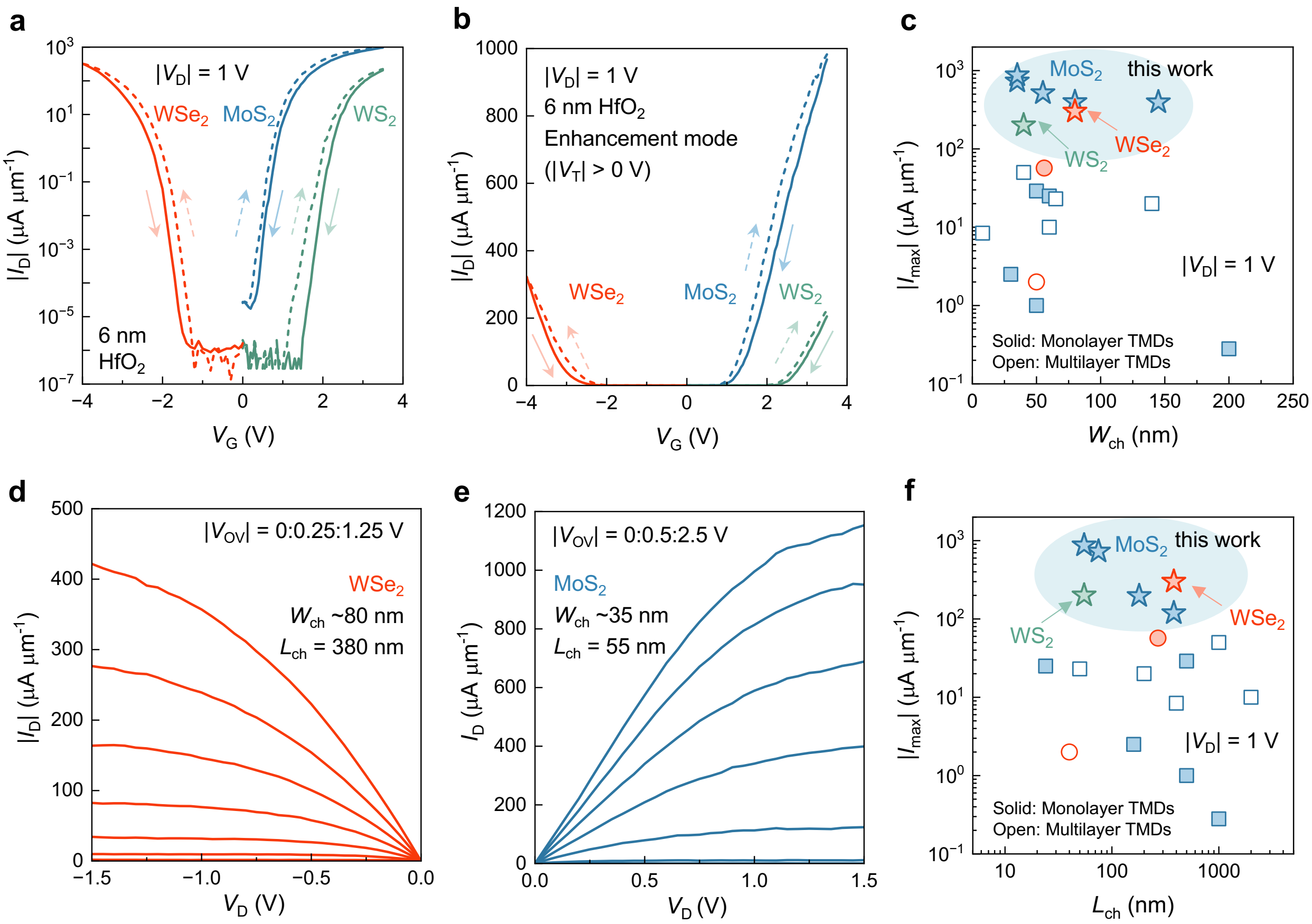

**Figure 4. High-performance complementary nanoribbon transistor based on monolayer TMDs. (a)** Log-scale and **(b)** linear-scale transfer characteristics ($I_D$-$V_{GS}$) of nanoribbon FETs based on $WSe_2$, $MoS_2$, and $WS_2$, demonstrating complementary *p*-type and *n*-type operation with steep switching and high on-state current. The 6 nm $HfO_2$ dielectric grown at 200 °C reduces gate leakage compared with the 3 nm $HfO_2$ dielectric grown at 90 °C (**Fig. 2-3**). Benchmarking of maximum current density ($I_{max}$) versus **(c)** channel width ($W_{ch}$) and **(f)** channel length ($L_{ch}$) for TMD nanoribbon FETs against previously reported monolayer TMD devices, highlighting the superior performance of nanoribbon devices in this work. **(d)** Output characteristics ($I_D$-$V_{DS}$) of NO-doped monolayer $WSe_2$ nanoribbon p-FETs, achieving $I_{max}$ of ~357 μA μm$^{-1}$. **(e)** Output characteristics ($I_D$-$V_{DS}$) of monolayer $MoS_2$ nanoribbon n-FETs, achieving $I_{max}$ of ~995 μA μm$^{-1}$.

To demonstrate the versatility and robustness of channel-width scaling across different TMD materials, we fabricated high-performance *p*-type $WSe_2$, *n*-type $MoS_2$, and *n*-type $WS_2$ nanoribbon transistors following the same fabrication process described in the previous section, with one key modification to the gate dielectric.

Unlike the thin gate dielectric adopted in previous sections (3 nm $HfO_2$ grown at 90 °C), a 6 nm $HfO_2$ dielectric grown at 200 °C was employed to improve film quality, increase the dielectric constant, and reduce gate leakage[43]. In addition, nitric oxide (NO) doping was applied to $WSe_2$ devices to significantly enhance hole injection and reduce contact resistance[21,25,43,44]. We have found that the yield of nanoribbon devices for $WSe_2$ is lower, so we designed the $WSe_2$ width to be approximately 80 nm, compared to about 35 nm for $MoS_2$ and $WS_2$. **Figure 4a-b** shows log-scale and linear-scale transfer characteristics of $WSe_2$, $MoS_2$, and $WS_2$ nanoribbon FETs on the same axes, confirming complementary *p*-type and *n*-type operation with steep switching and high on-state current across all three materials. The channel-length dependence of $WS_2$ and $WSe_2$ nanoribbon devices is provided in **Supplementary Figs. 8 and 9**, respectively, confirming maintained electrostatic control across channel lengths from 55 nm to 780 nm.

The NO-doped $WSe_2$ nanoribbon p-FETs (**Fig. 4d**) exhibit well-saturated output characteristics for a channel length of 380 nm, reaching a maximum current density ($I_{max}$) of ~357 μA μm$^{-1}$. Despite the enhancement provided by NO doping, threshold voltage ($V_{T,lin}$ = -3 V) remains quite negative, highlighting the need for additional $V_T$ tuning at aggressively scaled dimensions. The $MoS_2$ nanoribbon n-FETs (**Fig. 4e**) demonstrate $I_{max}$ of ~995 μA μm$^{-1}$ at $V_{OV}$ = 2.5 V with $V_{T,lin}$ = 1 V. The $WS_2$ nanoribbon n-FETs have a more positive $V_T$ than $MoS_2$ devices, attributable to differences in electron affinity relative to the gate metal and lower sulfur vacancy density (**Fig. 4a-b**). Note that all TMD channels explored here have reached enhancement mode operation ($|V_{T,lin}|$ > 0 V), which is important for low-power logic applications. These results are benchmarked against previously reported monolayer TMD nanoribbon and planar FETs in **Fig. 4c** and **Fig. 4f**, demonstrating that the nanoribbon devices in this work surpass prior reports across all three TMD materials. Summary of reported TMD nanoribbon transistors can be found in earlier work[36]. Hysteresis characteristics for $MoS_2$, $WS_2$, and $WSe_2$ nanoribbon devices are provided in **Supplementary Fig. 10**, showing small hysteresis of ~20 mV, ~90 mV, and ~15 mV, respectively, under standard measurement conditions without post-fabrication annealing.

## Extremely scaled monolayer nanoribbon FETs

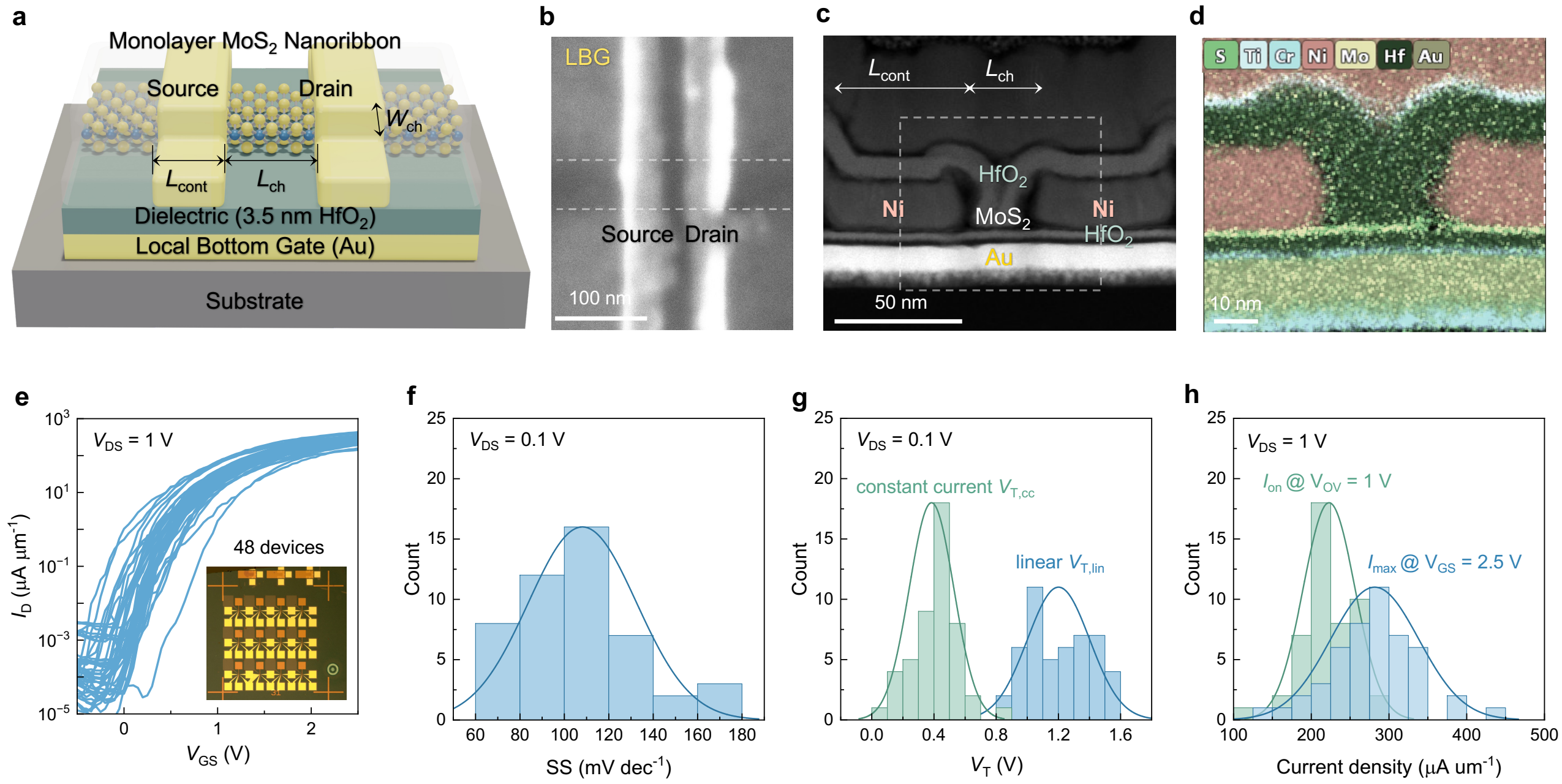

**Figure 5. Ultra-scaled monolayer $MoS_2$ nanoribbon FETs. (a)** Three-dimensional device schematic showing the ultrascaled $MoS_2$ nanoribbon transistors with contact length $L_{cont}$, channel length $L_{ch}$, and channel width $W_{ch}$, and 3.5 nm $HfO_2$ gate dielectric grown at 200 °C (EOT ~0.9 nm) on a local bottom gate. **(b)** SEM image of an ultrascaled nanoribbon transistor with $W_{ch}$ ~80 nm, $L_{ch}$ ~30 nm, and $L_{cont}$ ~30 nm. **(c)** Cross-sectional STEM image of a representative device with $L_{cont}$ = 50 nm and $L_{ch}$ = 30 nm. Scale bar, 50 nm. The top $HfO_2$ layer is deposited only during TEM testing and is not used for top-gating electrical transport. **(d)** The EELS maps illustrate the spatial distributions of the elements of the transistors. **(e)** Statistical $I_D$-$V_{GS}$ curves of 48 devices demonstrating consistent switching behavior and low device-to-device variability. Inset is an optical microscope image of ultrascaled $MoS_2$ nanoribbon transistor arrays. Scale bar, 200 nm. Histogram plots of **(f)** SS, **(g)** $V_T$, and **(h)** current density from 48 devices, showing mean SS = 108 mV dec$^{-1}$ (σ = 24 mV dec$^{-1}$), mean $V_{T,cc}$ = 0.38 V (σ = 0.14 V), and mean $I_{on}$ = 223 μA μm$^{-1}$ (σ = 33 μA μm$^{-1}$) at $V_{DS}$ = 1 V and $V_{OV}$ = 1 V.

Beyond width scaling, achieving ultimate scalability requires the scaling of channel length ($L_{ch}$), contact length ($L_{cont}$), and gate oxide thickness. Here, we fabricated ultra-scaled monolayer $MoS_2$ nanoribbon FETs with $W_{ch}$ of ~80 nm, $L_{ch}$ of ~30 nm, $L_{cont}$ = 70, 50, 30 nm, and 3.5 nm $HfO_2$ gate dielectric (EOT ~0.9 nm), with a schematic shown in **Fig. 5a**. SEM image of an ultrascaled nanoribbon transistor with $W_{ch}$ ~80 nm, $L_{ch}$ ~30 nm, and $L_{cont}$ ~30 nm is shown in **Fig. 5b**. The device geometry and gate stack are directly confirmed by high-angle annular dark-field scanning transmission electron microscope (STEM) in **Fig. 5c** and electron energy-loss spectroscopy (EELS) map in **Fig. 5d**, which resolve the 3.5 nm $HfO_2$ dielectric, the local bottom gate, and 50-nm $L_{cont}$ and 30-nm $L_{ch}$. Note that the top $HfO_2$ layer is deposited only during TEM testing and is not used for top-gating electrical transport. To characterize the dielectric stack, we performed metal–insulator–metal (MIM) capacitor measurements on 7 nm $HfO_2$ films to extract the equivalent oxide thickness (EOT), as shown in **Supplementary Fig. 11**. The dielectric constant is extracted to be approximately 15. With this value, the EOT of 3.5 nm $HfO_2$ is estimated to be around 0.9 nm for devices shown in **Fig. 5**. A similar gate stack has been reported in our previous works[43–45]. Statistical analysis of 48 ultrascaled devices (**Fig. 5e-h**) reveals a mean SS of ~107 mV $dec^{-1}$ (standard deviation 23 mV $dec^{-1}$) and a mean $V_{T,cc}$ of ~0.11 V (standard deviation 0.14 V). This variability still lags behind Si transistors, motivating a systematic study to understand the source of variability and how to reduce it through further oxide thinning, a double-gate architecture, and defect engineering. Further reduction of contact resistance through semimetal contacts[16,46] would enable continued contact length scaling, but it lies beyond the scope of this study.

## Discussion

This work demonstrates that channel-width scaling in monolayer TMD nanoribbon transistors and establish width as a critical design parameter for future 2D transistor technologies. Within the experimentally accessible regime (down to 35 nm), we observe continuous improvement in subthreshold swing (SS) and on-current ($I_{on}$), which we attribute to enhanced electrostatic control and contact injection. At more aggressive scaling (<10 nm), however, edge and roughness scattering are expected to become increasingly significant, potentially offsetting these benefits. This suggests the existence of an optimal channel width below ~30 nm, which depends on the transport regime and application requirements. From a device-technology co-optimization (DTCO) perspective, the optimal width is application-specific: ultra-narrow ribbons may be preferred for high-density logic and SRAM to meet density requirements, while wider ribbons may be advantageous for high-performance logic applications.

Creating sub-10 nm nanoribbons with smooth edges, however, will require precise patterning through optimized lithography and high-selectivity etching processes. In this regime, effective passivation of edge states will be crucial to maintain carrier transport and reduce variability and disorder caused by potential inhomogeneity. Further research on carrier transport in sub-10-nm nanoribbons in the disorder-dominated regime, especially as a function of temperature and carrier density, will be crucial for determining the ultimate scaling limits.

# Methods

## Materials

2-inch MBE-grown monolayer $MoS_2$ and $WS_2$ films were purchased from 2D Semiconductor Co., Ltd.

2-inch CVD-grown monolayer $MoS_2$ and $WSe_2$ films were provided by Nexstrom Pte. Ltd.

## Raman characterization

Raman spectroscopy was performed using a Thermo Scientific DXR3xi Raman Imaging Microscope with a 532 nm green laser, producing a spot size of approximately 1 μm. Each Raman mapping was conducted over ~10 $\mu m^2$ on a CVD-grown 1L-$MoS_2$ film from 2D Semiconductor Co., Ltd.

### Photoluminescence characterization

Two separate measurements were performed each using an alpha300R WitecSpectral imaging system. First, PL images were acquired on the edge of the monolayer $MoS_2$ test structures using 1 mW of 532 nm light focused on an approximately diffraction limited spot-size with a 100X/0.95 NA objective in a backscattering arrangement. Scattered light was dispersed with a Czerny-Turner spectrometer using a 300 L/mm grating resulting in a spectral accuracy of < 1.5 $cm^{-1}$ (<0.1 nm or <5 meV). Spectral were collected over a 5x5 μm area with acquisitions every 25 nm.

## Device fabrication

First, we pattern a Cr (3 nm) / Au (12 nm) local bottom gate (LBG), followed by ALD deposition of 3 nm ALD $HfO_2$ at 90°C ($\kappa$ ~10) for $MoS_2$ nanoribbon,[17] 6 nm $HfO_2$ at 200°C ($\kappa$ ~15) for $WS_2$ and $WSe_2$ nanoribbon,[43] and 3.5 nm $HfO_2$ at 200°C ($\kappa$ ~15) for extremely scaled $MoS_2$ devices, which serves as the bottom gate dielectric. Subsequently, wafer-grown 1L-$MoS_2$ films are wet transferred onto the LBG substrate. All nanoribbon channels with different $W_{ch}$ were patterned by electron-beam lithography (EBL). PMMA A4 resist was spin-coated at 6000 rpm for 60 s and baked at 180 °C for 5 min. The exposure base dose was 600 μC $cm^{-2}$ with proximity-effect correction. After exposure, the resist was developed in an IPA:$H_2O$ (3:1) solution at room temperature for 60 s, followed by a 20 s rinse in IPA. After EBL patterning, the channel nanoribbons were etched using an inductively coupled plasma (ICP) etch based on a $Cl_2/O_2$ chemistry. The detailed recipe is summarized as follows: $Cl_2$ flow = 15 sccm, $O_2$ flow = 5 sccm, chamber pressure ≈ 3 Pa, RF source power = 40 W, bias power

= 40 W, and etch time = 15-30 s. Then, PMMA was removed by soaking in acetone at 50 °C for 10 minutes, followed by a rinse in IPA. Ni source/drain (S/D) contacts with different $L_{ch}$, and $L_{cont}$ are patterned using e-beam lithography and e-beam evaporation.

### Nanoribbon width extraction

The nanoribbon channel width ($W_{ch}$) is extracted from SEM images using a quantitative and reproducible image-analysis procedure, as shown in **Supplementary Figs. 2**. Specifically, $W_{ch}$ is determined from SEM contrast line profiles taken perpendicular to the nanoribbon direction. The profiles are extracted using ImageJ and averaged over multiple positions along the channel to minimize local noise and edge roughness. The width is then defined as the full-width at half-maximum (FWHM) obtained from Gaussian fitting (**Supplementary Fig. 2**, bottom panels), providing a consistent definition of the effective channel width. Using this method, we estimate an uncertainty of ~3-5 nm, arising from SEM resolution, edge roughness, and fitting variability. Spatial variation is evaluated by sampling multiple locations along the channel, showing only minor variation and thus good width uniformity. Representative SEM images (**Supplementary Figs. 1 and 2**) show nanoribbons with widths precisely defined from ~35 nm to ~145 nm, while AFM characterization (**Supplementary Fig. 3**) further confirms the nanoscale width and structural continuity.

### Electrical characterization

The electrical characterization under vacuum ($\sim 1\times10^{-5}$ torr) was performed using a Keysight 4156C Precision Semiconductor Parameter Analyzer.

## Data Availability

Relevant data supporting the key findings of this study are available within the article and the Supplementary Information file. All raw data generated during the current study are available from the corresponding authors upon request.

## References

1. Chenming Hu *et al.* FinFET-a self-aligned double-gate MOSFET scalable to 20 nm. *IEEE Trans. Electron Devices* **47**, 2320–2325 (2000).

2. Colinge, J.-P. *et al.* Nanowire transistors without junctions. *Nat. Nanotechnol.* **5**, 225–229 (2010).

3. Loubet, N. *et al.* Stacked nanosheet gate-all-around transistor to enable scaling beyond FinFET. in *2017 Symposium on VLSI Technology* T230–T231 (IEEE, Kyoto, Japan, 2017). doi:10.23919/VLSIT.2017.7998183.

4. Agrawal, A. *et al.* Silicon RibbonFET CMOS at 6nm Gate Length. in *2024 IEEE International Electron Devices Meeting (IEDM)* 1–4 (2024). doi:10.1109/IEDM50854.2024.10873367.

5. Huang, C. Y. *et al.* 3-D Self-aligned Stacked NMOS-on-PMOS Nanoribbon Transistors for Continued Moore's Law Scaling. in *2020 IEEE International Electron Devices Meeting (IEDM)* 20.6.1-20.6.4 (IEEE, San Francisco, CA, USA, 2020). doi:10.1109/IEDM13553.2020.9372066.

6. Chhowalla, M., Jena, D. & Zhang, H. Two-dimensional semiconductors for transistors. *Nat. Rev. Mater.* **1**, 16052 (2016).

7. Akinwande, D. *et al.* Graphene and two-dimensional materials for silicon technology. *Nature* **573**, 507–518 (2019).

8. Liu, C. *et al.* Two-dimensional materials for next-generation computing technologies. *Nat. Nanotechnol.* **15**, 545–557 (2020).

9. Liu, Y. *et al.* Promises and prospects of two-dimensional transistors. *Nature* **591**, 43–53 (2021).

10. Zhu, K. *et al.* The development of integrated circuits based on two-dimensional materials. *Nat. Electron.* **4**, 775–785 (2021).

11. Das, S. *et al.* Transistors based on two-dimensional materials for future integrated circuits. *Nat. Electron.* **4**, 786–799 (2021).

12. Jiang, J., Xu, L., Qiu, C. & Peng, L.-M. Ballistic two-dimensional InSe transistors. *Nature* **616**, 470–475 (2023).

13. Desai, S. B. *et al.* $MoS_2$ transistors with 1-nanometer gate lengths. *Science* **354**, 99–102 (2016).

14. Shen, P. C. *et al.* Ultralow contact resistance between semimetal and monolayer semiconductors. *Nature* **593**, 211–217 (2021).

15. Lan, H.-Y., Appenzeller, J. & Chen, Z. Dielectric Interface Engineering for High-Performance Monolayer $MoS_2$ Transistors via hBN Interfacial Layer and Ta Seeding. in *2022 International Electron Devices Meeting (IEDM)* 7.7.1-7.7.4 (IEEE, San Francisco, CA, USA, 2022). doi:10.1109/IEDM45625.2022.10019439.

16. Li, W. *et al.* Approaching the quantum limit in two-dimensional semiconductor contacts. *Nature* **613**, 274–279 (2023).

17. Lan, H.-Y., Oleshko, V. P., Davydov, A. V., Appenzeller, J. & Chen, Z. Dielectric Interface Engineering for High-Performance Monolayer $MoS_2$ Transistors via $TaO_x$ Interfacial Layer. *IEEE Trans. Electron Devices* **70**, 2067–2074 (2023).

18. Jiang, J. *et al.* Yttrium-doping-induced metallization of molybdenum disulfide for ohmic contacts in two-dimensional transistors. *Nat. Electron.* **7**, 545–556 (2024).

19. Sun, Z. *et al.* Low Contact Resistance on Monolayer $MoS_2$ Field-Effect Transistors Achieved by CMOS-Compatible Metal Contacts. *ACS Nano* **18**, 22444–22453 (2024).

20. O'Brien, K. P. *et al.* Advancing 2D Monolayer CMOS Through Contact, Channel and Interface Engineering. in *2021 IEEE International Electron Devices Meeting (IEDM)* 7.1.1-7.1.4 (IEEE, San Francisco, CA, USA, 2021). doi:10.1109/IEDM19574.2021.9720651.

21. Chiang, C.-C., Lan, H.-Y., Pang, C.-S., Appenzeller, J. & Chen, Z. Air-Stable P-Doping in Record High-Performance Monolayer $WSe_2$ Devices. *IEEE Electron Device Lett.* **43**, 319–322 (2022).

22. Chou, A.-S. *et al.* High-Performance Monolayer $WSe_2$ p/n FETs via Antimony-Platinum Modulated Contact Technology towards 2D CMOS Electronics. in *2022 International Electron Devices Meeting (IEDM)* 7.2.1-7.2.4 (IEEE, San Francisco, CA, USA, 2022). doi:10.1109/IEDM45625.2022.10019491.

23. Wu, R. *et al.* Bilayer tungsten diselenide transistors with on-state currents exceeding 1.5 milliamperes per micrometre. *Nat. Electron.* **5**, 497–504 (2022).

24. Xiong, X. *et al.* Top-Gate CVD $WSe_2$ pFETs with Record-High $I_d$ ~594 μA/μm, $G_m$ ~244 μS/μm and $WSe_2/MoS_2$ CFET based Half-adder Circuit Using Monolithic 3D Integration. in *2022 International Electron Devices Meeting (IEDM)* 20.6.1-20.6.4 (IEEE, San Francisco, CA, USA, 2022). doi:10.1109/IEDM45625.2022.10019476.

25. Lan, H.-Y., Tripathi, R., Liu, X., Appenzeller, J. & Chen, Z. Wafer-scale CVD Monolayer $WSe_2$ p-FETs with Record-high 727 μA/μm $I_{ON}$ and 490 μS/μm $g_{max}$ via Hybrid Charge Transfer and Molecular Doping.

in *2023 International Electron Devices Meeting (IEDM)* 1–4 (IEEE, San Francisco, CA, USA, 2023). doi:10.1109/IEDM45741.2023.10413736.

26. Lan, H.-Y. *et al.* Stable Nitric Oxide Doping in Monolayer WSe2 for High-Performance P-type Transistors. Preprint at https://doi.org/10.21203/rs.3.rs-4916442/v1 (2024).

27. Mortelmans, W. *et al.* Record Performance in GAA 2D NMOS and PMOS Using Monolayer MoS2 and WSe2 with Scaled Contact and Gate Length. in *2024 IEEE Symposium on VLSI Technology and Circuits (VLSI Technology and Circuits)* 1–2 (IEEE, Honolulu, HI, USA, 2024). doi:10.1109/VLSITechnologyandCir46783.2024.10631395.

28. Pal, A., Bazizi, E. M., Colombeau, B., Alexander, B. & Ayyagari-Sangamalli, B. Nanosheet Width Investigation for Gate-All-Around Devices Targeting SRAM Application. in *2021 International Conference on Simulation of Semiconductor Processes and Devices (SISPAD)* 19–22 (IEEE, Dallas, TX, USA, 2021). doi:10.1109/SISPAD54002.2021.9592579.

29. IEEE International Roadmap for Devices and Systems 2024 Update More Moore, https://irds.ieee.org/.

30. Yinxiao, Y. & Murali, R. Impact of Size Effect on Graphene Nanoribbon Transport. *IEEE Electron Device Lett.* **31**, 237–239 (2010).

31. Arutchelvan, G. *et al.* Impact of device scaling on the electrical properties of $MoS_2$ field-effect transistors. *Sci. Rep.* **11**, 6610 (2021).

32. Zhang, Z. *et al.* A Gate-All-Around inO Nanoribbon FET With Near 20 mA/m Drain Current <sub/> <sub/> <sub/>. *IEEE Electron Device Lett.* **43**, 1905–1908 (2022).

33. Chen, S. *et al.* Monolayer MoS2 Nanoribbon Transistors Fabricated by Scanning Probe Lithography. *Nano Lett* **19**, 2092–2098 (2019).

34. Kotekar-Patil, D., Deng, J., Wong, S. L., Lau, C. S. & Goh, K. E. J. Single layer MoS2 nanoribbon field effect transistor. *Appl. Phys. Lett.* **114**, (2019).

35. Duan, X. *et al.* $MoS_2$ Nanoribbon Transistor for Logic Electronics. *IEEE Trans. Electron Devices* 1–6 (2022) doi:10.1109/TED.2022.3164859.

36. Chen, S., Zhang, Y., King, W. P., Bashir, R. & Van Der Zande, A. M. Edge-Passivated Monolayer $WSe_2$ Nanoribbon Transistors. *Adv. Mater.* 2313694 (2024) doi:10.1002/adma.202313694.

37. Hoque, Md. A. *et al.* Ultranarrow Semiconductor WS2 Nanoribbon Field-Effect Transistors. *Nano Lett.* **25**, 1750–1757 (2025).

38. Panarella, L. *et al.* Evidence of contact-induced variability in industrially-fabricated highly-scaled MoS2 FETs. *Npj 2D Mater. Appl.* **8**, 44 (2024).

39. Wang, K., Taniguchi, T., Watanabe, K. & Xue, J. Natural p–n Junctions at the $MoS_2$ Flake Edges. *ACS Appl. Mater. Interfaces* **14**, 39039–39045 (2022).

40. Tang, J. *et al.* In Situ Oxygen Doping of Monolayer MoS2 for Novel Electronics. *Small* **16**, 2004276 (2020).

41. Shen, P.-C. *et al.* Healing of donor defect states in monolayer molybdenum disulfide using oxygen-incorporated chemical vapour deposition. *Nat. Electron.* https://doi.org/10.1038/s41928-021-00685-8 (2021) doi:10.1038/s41928-021-00685-8.

42. Eppler, D., Lan, H.-Y., Chen, Z. & Appenzeller, J. Investigation into Side Contact Behavior and Transfer Length of Monolayer MoS2 FETs. in *2025 IEEE International Electron Devices Meeting (IEDM)* 1–4 (2025). doi:10.1109/IEDM50572.2025.11353744.

43. Lan, H.-Y. *et al.* Uncovering the doping mechanism of nitric oxide in high-performance P-type $WSe_2$ transistors. *Nat. Commun.* **16**, 4160 (2025).

44. Lan, H.-Y. *et al.* Improved Hysteresis of High-Performance p-Type $WSe_2$ Transistors with Native Oxide $WO_x$ Interfacial Layer. *Nano Lett.* **25**, 5616–5623 (2025).

45. Lan, H.-Y. *et al.* Reliability of high-performance monolayer $MoS_2$ transistors on scaled high-κ $HfO_2$. *Npj 2D Mater. Appl.* **9**, 5 (2025).

46. Wu, W.-C. *et al.* On the Extreme Scaling of Transistors with Monolayer $MoS_2$ Channel. in *2024 IEEE Symposium on VLSI Technology and Circuits (VLSI Technology and Circuits)* 1–2 (IEEE, Honolulu, HI, USA, 2024). doi:10.1109/VLSITechnologyandCir46783.2024.10631401.

## Acknowledgements

This work was supported in part by the Semiconductor Research Corporation (SRC) and National Institute of Standards and Technology (NIST) through the NEW LIMITS Center under Award 70NANB17H041. This work was conducted using the facilities at the Birck Nanotechnology Center. We acknowledge I-Chen Tseng's support for visualizing device schematics.

## Author Contributions

H.-Y. L. proposed the original idea. Z. C. and J. A. supervised the project. H.-Y. L. performed device fabrication, characterization, simulation, and data analysis with the input from S.-H. Y., Y. Cho, J. Cai, and Z. Sun. S.-H.

Y. and Y. Tan performed STEM and SEM characterization. Y. Cho performed AFM characterization. C. Li. and L.-Y. H. grew the 2D materials with the input from Y. Wan and L.-J. L.. T. B. performed photoluminescence and data analysis. H.-Y. L. wrote the first draft of the manuscript with the input from T. B. and Z. C., and all authors discussed and revised the final manuscript.

## Competing Interests

The authors declare no competing financial interests.

## Supplementary Figures